\documentclass[%
 reprint,
amsmath,amssymb,
aps,
]{revtex4-1}

\usepackage{graphicx}
\usepackage[squaren, cdot]{SIunits}
\usepackage{lineno}
\usepackage{hyperref}

\begin{document}

\title{Synchronization of atomic quantum systems in multi-site optical trapping potentials}
\author{Malte Schlosser}
\author{Jens Kruse}
\author{Gerhard Birkl}
\email{gerhard.birkl@physik.tu-darmstadt.de}
\affiliation{Institut f\"{u}r Angewandte Physik, Technische Universit\"{a}t Darmstadt, Schlossgartenstra\ss e 7, 64289 Darmstadt, Germany}

\date{\today}



\begin{abstract}
	
	Advanced quantum technologies, such as quantum simulation, computation, and metrology are thriving for the implementation of large-scale configurations of identical quantum systems. Sets of atoms and molecules have the advantage of having identical intrinsic properties but need to be placed in identical environments as well. In this work, we present a strong suppression of dephasing effects and a significant increase of the coherence time for ensembles of neutral atoms in arrays of optical traps. Compensation of the differential Stark shift caused by the optical trapping potential is achieved by the superposition of a second, near resonant light field to the far-detuned trapping light. The achieved synchronization of the coherent evolution is demonstrated by analyzing the hyperfine-state phase evolution of atomic ensembles of $^{85}${Rb} trapped in a two-dimensional array of dipole traps via Ramsey spectroscopy. The experimental method presented here does not require the existence of a so called magic wavelength and is expandable to other atomic species trapped in various dipole trap configurations of arbitrary wavelength. 
	
\end{abstract}

\maketitle

\begin{figure}
	\includegraphics[width=1 \linewidth]{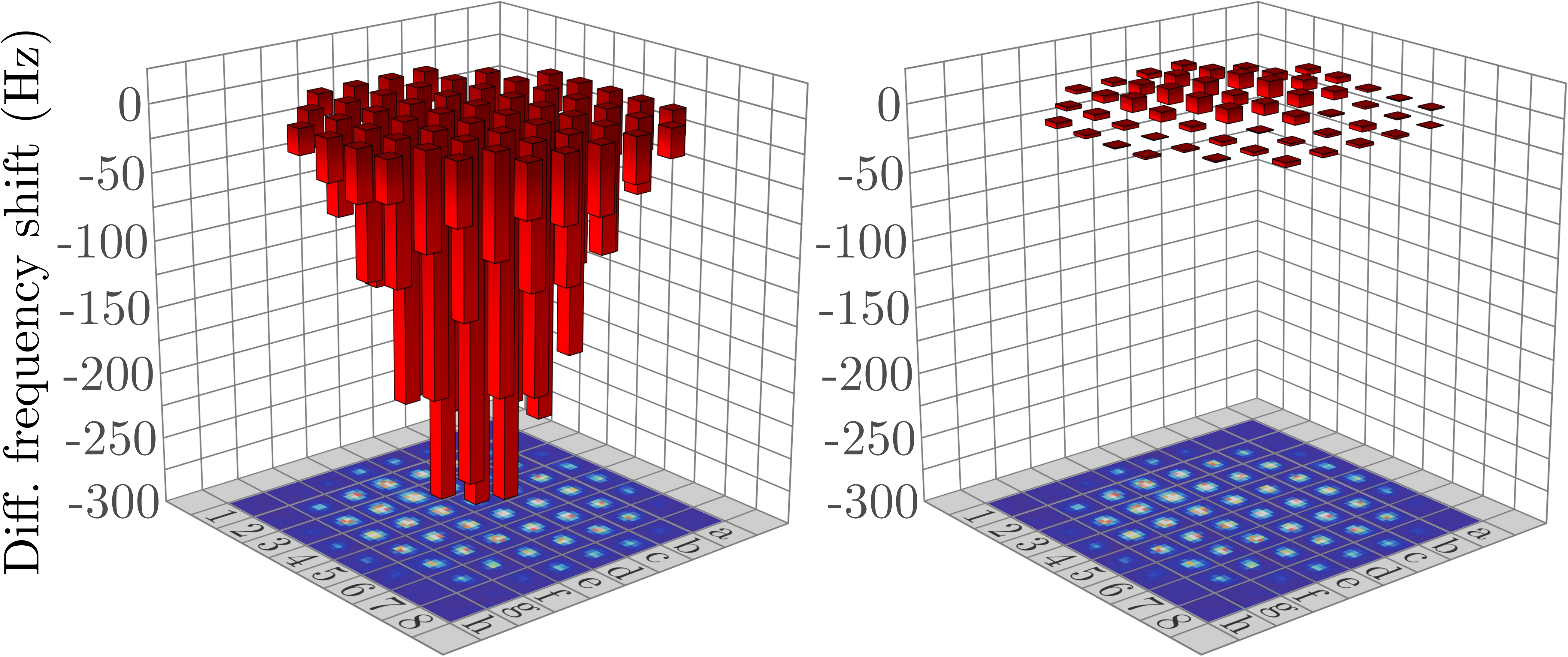}
	\caption{(Left) Measured differential frequency shift of the ground-state hyperfine splitting of $^{85}$Rb atoms in a two-dimensional array of dipole traps. A fluorescence image of the trapped atom ensembles is shown at the bottom. (Right) Residual frequency shift under identical trapping conditions with a weak additional laser field compensating for the differential light shift.
	\label{fig:two_d_array_comp}
	}
\end{figure}

%
%
In recent years there has been an enormous progress in the field of quantum engineering with ultracold neutral atoms trapped and manipulated by optical means. Applications in quantum metrology and quantum information science, such as all-optical atomic clocks or quantum information processing, benefit from the extraordinary extend of control provided by light fields with respect to the internal and external degrees of freedom of neutral atoms.
One main requirement in most of these experiments is the capability to precisely manipulate the electronic states of an atomic system which features long coherence times as well \cite{Ye_katori_kimble_LightTraps}. Unfortunately, in most quantum systems the influence of optical potentials is state dependent and thus perturbs a coherent evolution of quantum phases due to differential effects \cite{Beugnon_1, Schrader_QuantumRegister_PRL2004, Lundblad_field_sensitiv_adressing_porto, yavuz_Saffman:2006:PRL,lengwenus-2009}. Several approaches to separate internal and external dynamics have been made. Specially the realization of trapping atoms in potentials of a so called "magic wavelength" represents a break through in high precision spectroscopy like for example in optical clocks \cite{Katori_Sr_clock}.
A magic wavelength appears in some elements whose complex electronic structure results in the cancellation of differential effects for certain wavelengths of the trapping fields, thus yielding a vanishing perturbation of the atomic phase evolution. Although there are no nondisturbing optical potentials for alkali atoms generated by light of simple magic wavelength  \cite{Flammbaum_Micromagic_Clocks}, these elements still play a crucial role in quantum optics, representing a well determined bosonic test bed. One option to minimize the spatial dependence of the light field for alkalis is for example the trapping of an atomic ensemble in a far blue detuned dipole trap, where the atoms spend most of their time in a nondisturbed space \cite{Davidson_1}. Other possibilities are to use a vector potential of partly circular polarized light to compensate for differential light shifts \cite{Lundblad_Schlosser_Magic_wavelength} or to suppress dephasing mechanisms by a sequence of microwave pulses \cite{Davidson_2004_suppress_dephasing}.
\\
Here we present the experimental realization of a simple and universally extendable scheme of compensating the differential light shift induced by dipole traps. We perform Ramsey spectroscopy on $^{85}$Rb trapped in an array of dipole potentials to determine the coherence time. By using a second, near resonant light field we can compensate the differential light shift (Fig. \ref{fig:two_d_array_comp}) and strongly suppress atomic dephasing \cite{kaplan_davidson_suppresion,radnaev2010quantum,PhysRevLett.106.213002}. Due to this we increase the coherence time which is one of the most important key elements in quantum metrology and quantum information systems.
%
%
%
%
\\
The interaction of a multi-level atom with internal ground states $\mathrm{|g_i\rangle}$ and excited states $\mathrm{|e_j\rangle}$ and a laser field with frequency $\omega_L$ and intensity $\mathrm{I(\mathbf{r})}$ can be described to good approximation by the dipole interaction. In rotating-wave-approximation (RWA) the resulting ac-stark shift experienced by the atom in ground state $|g_i\rangle$ is given by    
\begin{equation}\label{dipolepotential}
	\Delta E_i(\mathbf{r})=\frac{3\pi c^2\Gamma}{2\omega_{ij}^3} \sum \frac{c_{ij}}{\Delta_{ij}} I(\mathbf{r})~.
\end{equation}
Here, $c$ are the speed of light and $\mathrm{\Gamma}$ the natural linewidth of the dipole transition, respectively. The summation takes into account the contributions of coupling to different excited states $\mathrm{|e_j\rangle}$, each with its respective interaction coefficient $\mathrm{c_{ij}}$, transition frequency $\mathrm{\omega_{ij}}$, and detuning $\mathrm{\Delta_{ij}=\omega_L-\omega_{ij}}$.
The energy shift can be used to generate a conservative trapping potential for the atoms known as dipole potential with a trap minimum at the maximum of $\mathrm{I(r)}$ for $\mathrm{\Delta_{ij} < 0}$.
\\
Most atoms like the alkali elements have more than one long-lived ground state due to their nuclear spin. Pairs of these hyperfine states, further referred to as $\mathrm{|0\rangle}$ and $\mathrm{|1\rangle}$, respectively, with energy difference $\mathrm{\hbar \omega_{HFS}}$
are central to applications, e.g. in metrology as frequency standards or as qubit basis in quantum computing and quantum simulation.
One drawback of optical trapping is the nonnegligible influence of the dipole trap on the energy difference of these pairs of ground states which yields a differential frequency shift
\begin{equation}
	 \delta_{diff}\left(\mathbf{r}\right)=\frac{3\pi c^2\Gamma}{2\hbar\omega_{ij}^3} \left(\sum \frac{c_{0j}}{\Delta_{0j}}-\sum \frac{c_{1j}}{\Delta_{1j}}\right)\cdot I(\mathbf{r})~.
\end{equation}
Even in the case of far-detuned optical traps based on linearly polarized light where the interaction coefficients for the two states can be considered equal $\mathrm{(c_{0j}=c_{1j})}$, the differential frequency shift caused by the difference in detuning $\mathrm{\Delta_{1j} = \Delta_{0j} + \omega_{HFS}}$ remains and causes a reduction of the effective hyperfine splitting for $\mathrm{\Delta_{ij} < 0}$.
This differential ac-Stark shift leads to a trap-depth-dependent shift of the frequency difference between $\mathrm{|0\rangle}$ and $\mathrm{|1\rangle}$ even for atoms occupying the trap ground state.
\\
The energy distribution of a thermal atom ensemble of finite temperature in a dipole trap causes a broadening with a probability density closely related to that of a three-dimensional Boltzmann distribution.
For coherent superpositions of internal states, this frequency broadening induces a dephasing of the atomic state evolution \cite{Kuhr_Meschede:2005:PRA} for atom ensembles but also for repeated realisations with single trapped atoms at finite temperature. 
For applications which rely on atoms being localized in multiple dipole traps, the differential frequency shift causes two major problems: 1) The effective resonance frequency of the atomic transition varies from trap to trap as a function of the specific trap depths. 2) The thermal distribution of energies causes an additional frequency spread within each trap.
As a consequence, the differential frequency shift limits the spectroscopic frequency resolution and the coherence times for atoms in optical traps.
\\
\begin{figure}
	\includegraphics[width=1 \linewidth]{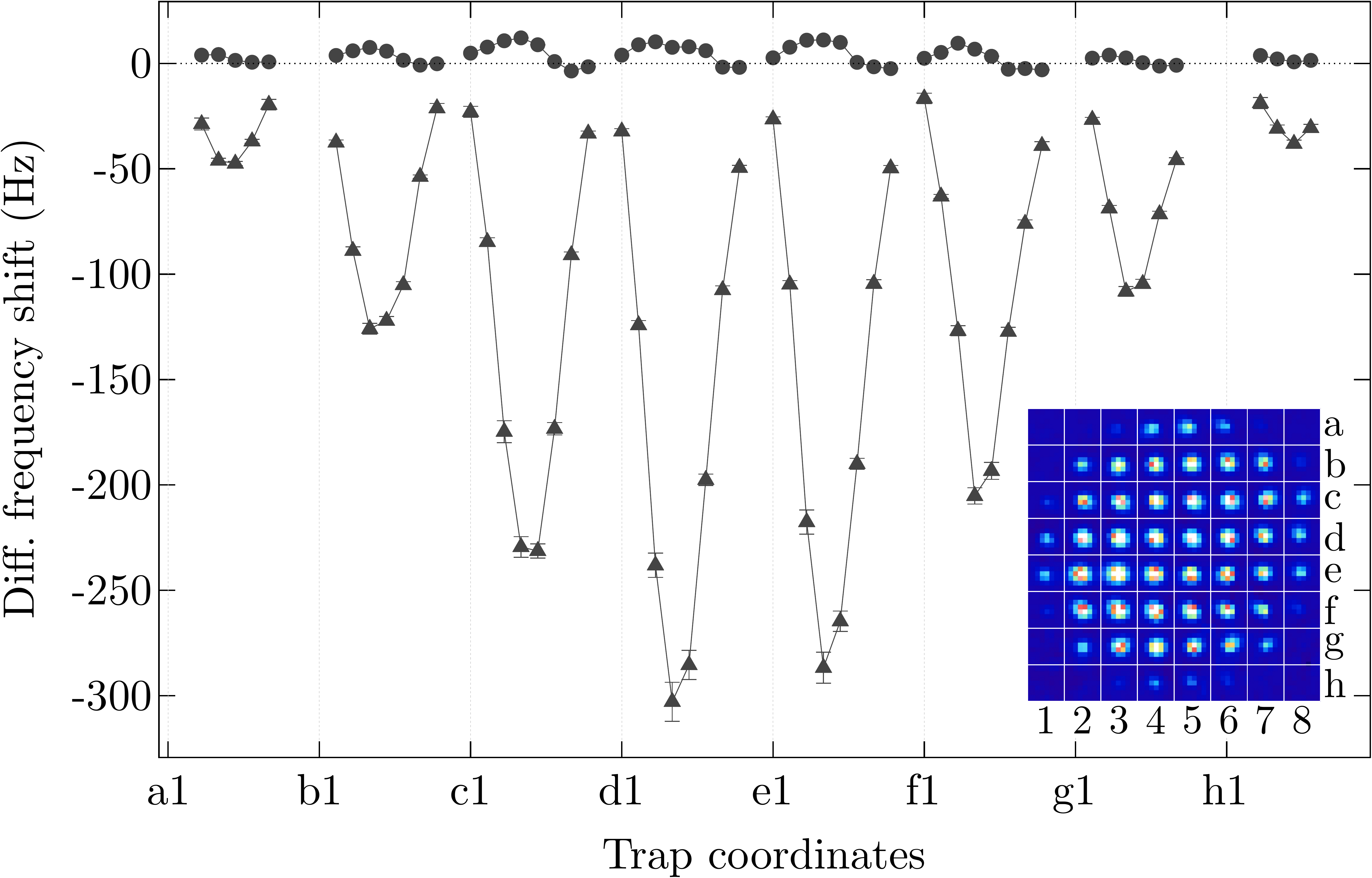}
	\caption{(Triangles) Differential frequency shift in a two-dimensional array of dipole traps measured by site-selective Ramsey spectroscopy on small ensembles of atoms. (Circles) Reduced frequency shift after compensation of the differential light shift by spatially overlapping an additional laser field detuned in between the two ground-state hyperfine levels. (Inset) Fluorescence image of the atom ensembles in the dipole trap array including the labeling of the set of traps.
		\label{fig:comp_freq}
	}
\end{figure}%
%
To compensate for the differential frequency shift, a second light field coupling the two hyperfine ground states to the excited state manifold in an appropriate fashion is introduced. The compensation field is detuned in between the two hyperfine ground states with $\mathrm{\Delta_{c,j}\approx-\omega_{HFS}/2}$ with respect to state $\mathrm{|0\rangle}$
yielding a differential shift of
\begin{equation}
	\delta_{c}\left( \mathbf{r}\right)=\frac{3\pi c^2\Gamma}{2\hbar\omega_{ij}^3} \left(\sum \frac{c_{0j}}{{\Delta_{c,j}}}+\sum \frac{c_{1j}}{{\omega_{HFS}+\Delta_{c,j}}}\right) I_{c}(\mathbf{r}).
\end{equation} 
The index ${j}$ in the detuning accounts for slight modifications caused by the frequency splitting of the hyperfine manifold of the excited state. Due to the change of sign in the detuning, the compensation laser field increases the effective hyperfine splitting.
Thus, by superimposing both light fields with identical intensity profiles one can compensate for the differential frequency shift and the ac stark shifts of both hyperfine ground states become identical. Only balancing of the differential shift is necessary, which typically is orders of magnitude smaller than the absolute ac stark shift \cite{r1}. As a result, the  change in the absolute trap depth is negligible. For compensation at every position $\mathrm{\mathbf{r}}$, the two light fields have to be perfectly mode matched 
\begin{equation}
	\delta_{diff}(\mathbf{r})+\delta_c(\mathbf{r})\stackrel{!}{=} 0\quad\rightarrow\quad I_c(\mathbf{r})=\eta \times I(\mathbf{r})
\end{equation} 
where the ratio $\mathrm{\eta}$ approximately scales with $\mathrm{\eta\approx \left(\Delta_{c,j}/\Delta_{ij}\right)^2 \approx
\left(\omega_{HFS}/(2\Delta_{ij})\right)^2}$ \cite{kaplan_davidson_suppresion, r2}.\\
Although the detuning $\mathrm{\Delta_{c,j}}$ is in the range of the ground state hyperfine splitting, the rates of spontaneous scattering of both laser fields ($\mathrm{\Gamma_{sc}}$ and $\mathrm{\Gamma_{sc,c}}$) are comparable with a ratio $\mathrm{\Gamma_{sc,c}/\Gamma_{sc} \approx 1}$ for typical trap laser detunings due to the small amount of required compensation intensity. 
Thus, for typical trap laser detunings, the compensation field only adds the same rate of spontaneous scattering events as the trapping light generates itself and the combined total rate of spontaneous scattering can be minimized by increasing the trap laser detuning.
\\
%
%
In the experimental implementation, we work with small ensembles of ultracold $^{85}$Rb atoms trapped in an array of well separated dipole traps (see inset of Fig. \ref{fig:comp_freq}). Our system represents a scalable architecture for quantum technology using the two hyperfine levels $\mathrm{|0\rangle = |F=2, m_F=0\rangle}$ and $\mathrm{|1\rangle = |F=3, m_F=0\rangle}$ ('clock states') as the qubit states \cite{Dumke_1,Schlosser2011,OhldeMello2019,Schlosser2019}. We perform single qubit rotations by coupling the two clock states via a coherent two photon process.
\\
The dipole trap array is created by illuminating a two-dimensional (2D) array of microlenses with a Gaussian laser beam \cite{Birkl_optics_communications}. The lenses of $\mathrm{\unit{50}{\mu m}}$ radius are arranged in a quadratic grid with a spacing of  $\mathrm{\unit{125}{\mu m}}$. By reimaging the resulting focal plane into a vacuum chamber we superimpose the dipole trap array onto a cold cloud of $^{85}$Rb atoms cooled in a MOT. We load each of the traps with several tens to hundreds of atoms during a sequence of optical molasses. All traps have the same waist of $\mathrm{w_{trap}=\unit{3.7\pm 0.1}{\mu m}}$ ($\mathrm{1/e^2}$-radius). The trap separation is \unit{58.6 \pm 1.6}{\mu m} after reimaging. 
The trapping light is derived from a Titanium Saphire laser (TiSa) at a wavelength of $\mathrm{\lambda_{TiSa}=\unit{810.1}{nm}}$. The light is intensity stabilized to $\mathrm{\Delta P/P\approx 10^{-4}}$ at a power of $\mathrm{P_{TiSa}=\unit{41\pm 1}{mW}}$. The Gaussian beam illuminating the microlens array has a beam waist of $\mathrm{\unit{500\pm 3}{\mu m}}$. This results in a power of $\mathrm{P= \unit{0.72}{mW}}$ in the central trap leading to a trap depth of $\mathrm{U=k_B\cdot \unit{40}{\mu K}}$. The trapping frequencies are $\mathrm{2\pi\cdot \unit{5.5}{ kHz}}$ in radial and $\mathrm{2\pi\cdot \unit{270}{Hz}}$ in axial direction. We have the ability to resolve all single traps in parallel with a CCD-camera by fluorescence imaging and we showed before that we can address single traps with a tight focused laser beam \cite{Dumke_1} or a spatial light modulator \cite{kruse-2010}.
\\
Site-resolved Ramsey spectroscopy and its extension to spin-echo spectroscopy allow us to determine the transition frequency and the coherence time of the superposition of the clock states in each trap. The coupling of the two hyperfine states is realized by a pair of phase-locked laser fields with intentional detuning $\mathrm{\delta_{RL}}$ relative to $\mathrm{\omega_{HFS}}$. From the measured frequency of the Ramsey oscillations we determine the shift of the atomic resonance frequency of the trapped atoms from the the unperturbed hyperfine splitting $\mathrm{\omega_{HFS}}$: $\mathrm{\omega_{meas}=\omega_{HFS}+\delta_{RL}}$. One contribution to the observed shift is the difference $\mathrm{\delta_{B}}$ of the second order magnetic field shifts $\mathrm{\delta_{B}=\unit{(264\pm 6)}{Hz}}$ of the two clock states. This offset occurs due to the applied magnetic bias field of $\mathrm{B=\unit{(45.2\pm 0.5)}{\mu T}}$, which is assumed to be identical for all traps. The remaining contribution to the frequency shift is the differential ac-Stark shift $\delta_{diff}$ induced by the trapping light. 
In Fig. \ref{fig:comp_freq} (triangles) the measured frequency shift 
$\mathrm{\delta_{diff} = \omega_{meas}-\omega_{HFS}-\delta_{B}}$
of atom ensembles trapped in 54 dipole traps is presented.
The observed differential light shift reflects the spatial intensity profile of the Gaussian dipole beam, which is evident in the representation of Fig. \ref{fig:two_d_array_comp} (left). The decreased frequency of deep center traps compared to the small shift of more shallow traps leads to a frequency spread of $\mathrm{\unit{(286.3\pm 11.7)}Hz}$ regarding all trapped atom ensembles.
\\
For the compensation of the differential ac-Stark shift a coupling of the clock states via the D1 or D2 transitions is feasible. However, to minimize variations in the spatial overlap due to chromatic aberrations, we chose a compensation light field close to resonance with the D1 transition. 
To implement the scheme, we use a second TiSa laser, frequency stabilized at a wavelength of $\mathrm{\lambda_{c}=\unit{794.978}{nm}}$ centered in between the ground state hyperfine levels. Both laser fields are spatially mode-matched by sending them through the same polarization maintaining single-mode optical fiber.
We determine the laser power $\mathrm{P_c}$ required for optimal compensation by experimentally minimizing the frequency spread between different dipole traps. In Fig. \ref{fig:comp_freq}  (circles) the frequency shift of the 54 analyzed atomic ensembles in the dipole trap array with the additional compensation field is presented. The frequency spread is reduced to 
$\mathrm{\unit{(15.7 \pm 0.3)}Hz}$. Compared to the differential light shift induced by the trapping beam, the frequency spread between the compensated traps is reduced by a factor of 18.2. 
The average value of the remaining frequency shift is $\mathrm{\unit{(3.5\pm 3.6)}Hz}$.
For this result, the compensation laser had a measured total power of $\mathrm{P_c=\unit{4.3\pm 0.2}{nW}}$ in front of the microlens array.
%
\begin{figure}
	\includegraphics[width=1\linewidth]{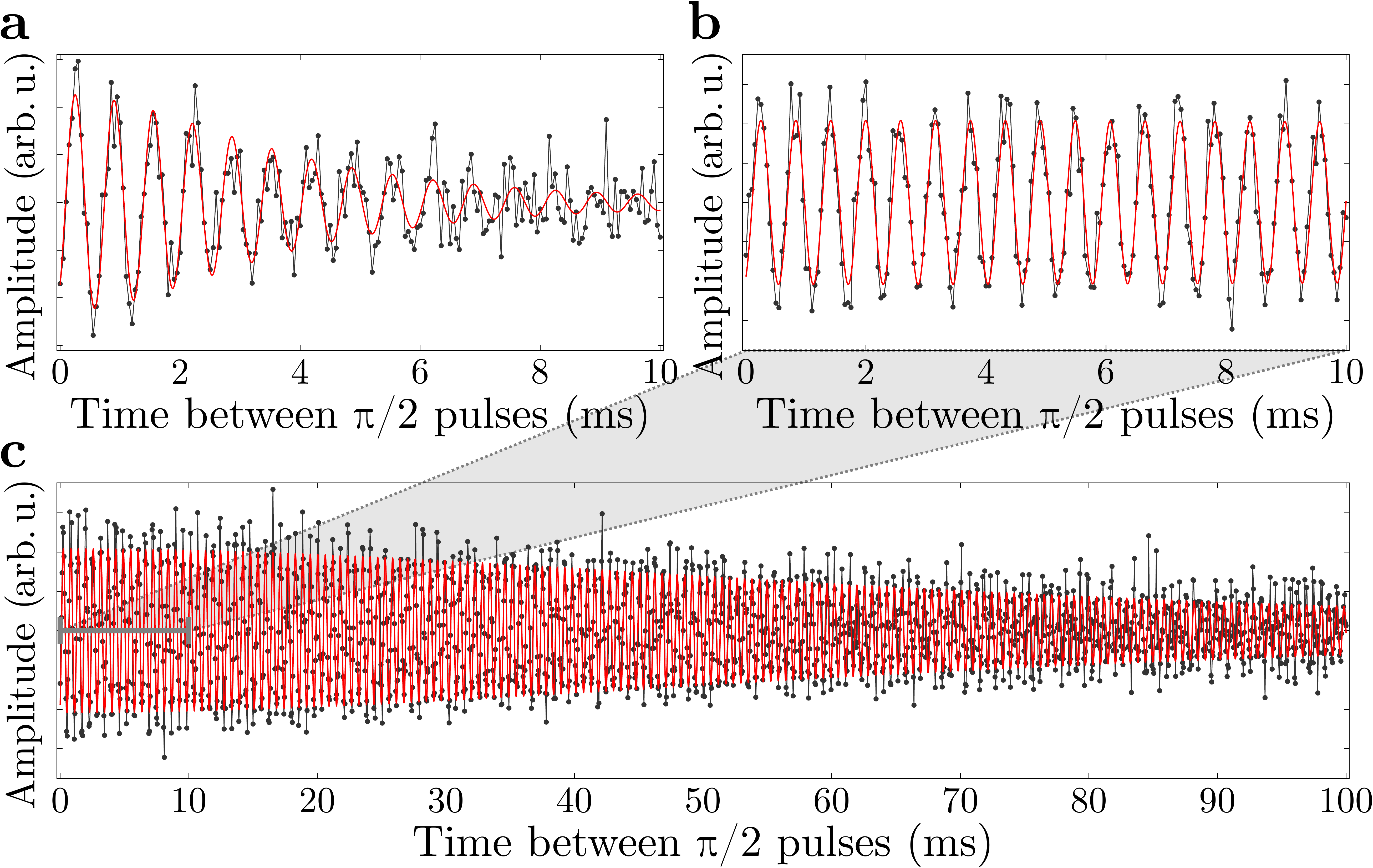}
	\caption{Ramsey oscillations of a thermal atom ensemble in a single central trap of the dipole trap array (trap labeled e5 in Figs. \ref{fig:two_d_array_comp} and \ref{fig:comp_freq}) (a) without compensation of the differential ac-Stark shift induced by the dipole trap and (b,c) with compensation. The dephasing time increases by a factor of 19.6 due to compensation.
		\label{fig:ramsey_single}
	}
\end{figure}%
%
%
\begin{figure}
	\includegraphics[width=1\linewidth]{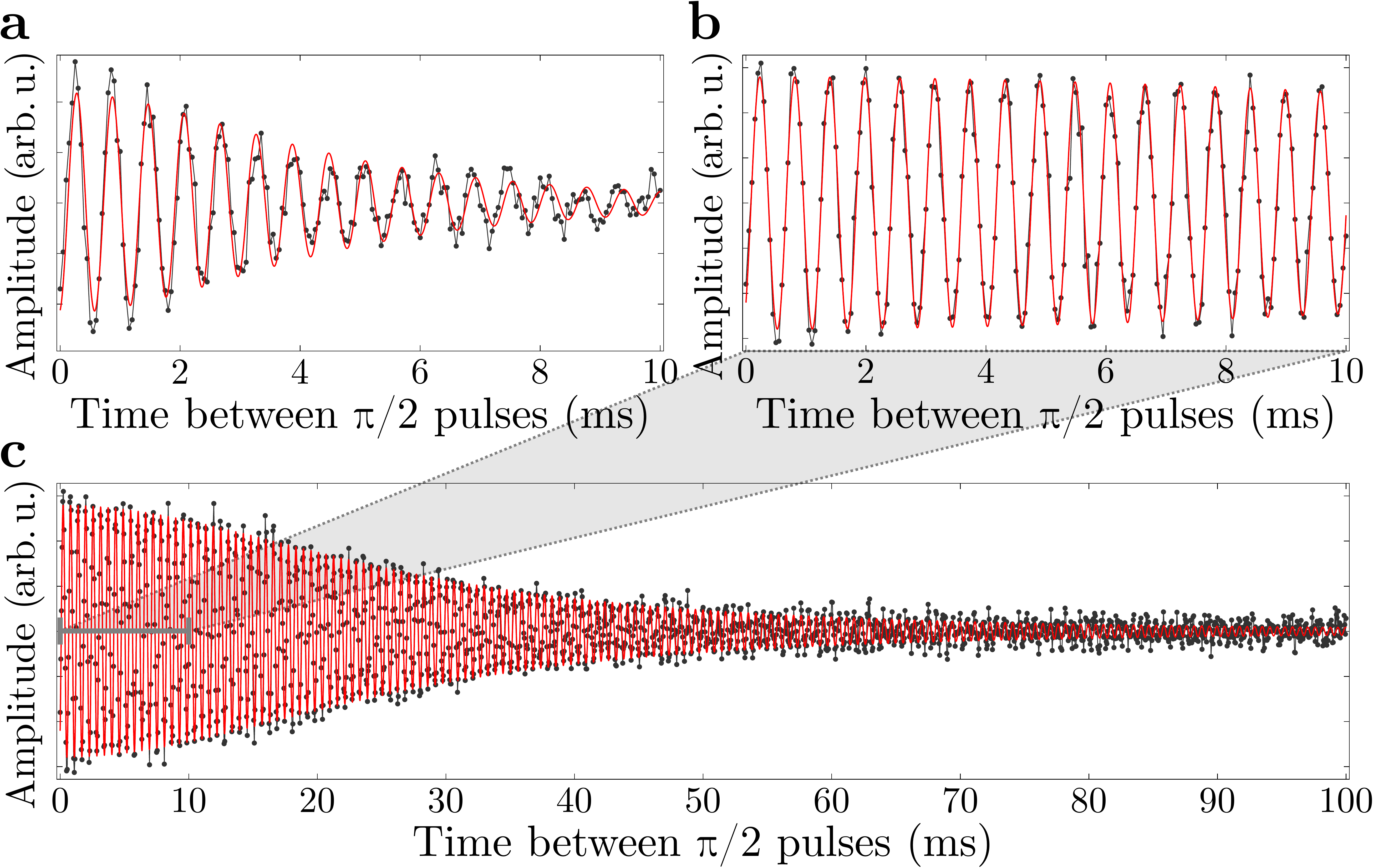}
	\caption{Ensemble average of the Ramsey oscillations of all 54 occupied traps (a) without compensation of the differential ac-Stark shift and (b,c) with compensation. The dephasing time for the whole ensemble increases by a factor of 6.2 due to compensation.
	\label{fig:ramsey_all}
	}
\end{figure}%
%
This results in a power ratio between both laser fields of $\mathrm{\eta_{meas} = 1.05\pm 0.07\times 10^{-7}}$.\\
We attribute the residual spatial structure remaining in the compensated frequency spread (see also Fig. \ref{fig:two_d_array_comp} (right)) to a slight displacement between the two laser fields of wavelength \unit{810}{nm} and \unit{795}{nm} at the position of the microlens array. By assuming a misalignment between both beams of \unit{8}{\mu m}, the residual structure in the spatial distribution of frequency spreads is matched. The displacement of the two beams can be caused by a slightly decentered (\unit{200}{\mu m}) collimation lens in the fiber output coupler used for both beams. Dispersive effects in the collimation lens cause the lateral displacement of the two beams impinging on the microlens array. Assuming a centered collimation lens, would reduce the residual frequency spread to $\mathrm{\unit{1.7}{Hz}}$ between all 54 dipole traps.
\\
The compensation field does not only cancel the differential light shift for trapped atoms and equalizes the observed hyperfine splitting in different dipole traps, it also strongly reduces the dephasing of the coherent evolution of atoms in a single trap as well of the ensemble average of all traps in the trap array. 
In Fig. \ref{fig:ramsey_single}, Ramsey oscillations of the population of the lower clock state are shown as a function of the time of free evolution for a single trap (trap labeled e5 in Figs. \ref{fig:two_d_array_comp} and \ref{fig:comp_freq}) (a) without and (b, c) with the compensation field.
The spread of the differential ac-Stark shifts causes a dephasing of the coherent evolution of different atoms which results in a temporal reduction of the amplitude of Ramsey oscillations which can be characterized by the inhomogeneous dephasing time $\mathrm{T^*_2}$ \cite{Lengwenus_1}. Without compensation, we observe a dephasing time $\mathrm{T^*_2 = \unit{(4.46 \pm 0.31)}{ms}}$ in Fig. \ref{fig:ramsey_single} (a). Compensation of the differential ac-Stark shift, strongly suppresses dephasing and increases the dephasing time to $\mathrm{T^*_2 = \unit{(87.5 \pm 2.1)}{ms}}$ (Fig. \ref{fig:ramsey_single} (b)). This represents an increase in the dephasing time by a factor of 19.6. 
After a free-evolution time of \unit{100}{ms}, Ramsey oscillations are still clearly observable.\\
We could also achieve a significant reduction of dephasing in the combined Ramsey signal of all 54 occupied traps. Figure \ref{fig:ramsey_all} presents the resulting data in the same way as Fig. \ref{fig:ramsey_single} for a single trap. Without compensation, the trap average gives a dephasing time of $\mathrm{T^*_2 = \unit{(5.21 \pm 0.25)}{ms}}$ which is increased by a factor of 6.2 to $\mathrm{T^*_2 = \unit{(32.4 \pm 3.3)}{ms}}$ when including the compensation field. As discussed before, in contrast to a specific trap, the perfect cancellation of the differential ac-Stark shift could not be achieved for all traps at the same value of $\mathrm{P_c}$ due to the lateral shift of the two laser beams. As a consequence, the factor for the increase of the dephasing time has to turn out smaller for the trap ensemble than for a single trap.
\\
In order to determine the coherence time, here given by the time constant of irreversible dephasing and spontaneous scattering, we extend the Ramsey to a spin-echo measurement by applying a state-inverting $\pi$-pulse applied after the first half of the free-evolution time. Due to this technique, any dephasing caused by static inhomogeneities in the first half of the free evolution is reversed in the second half and thus is not leading to a decay in the oscillation amplitude of the state population in the spin-echo signal. This makes this technique sensitive to irreversible dephasing mechanisms. In our situation, irreversible dephasing is mainly caused by atom heating, leading to a change in the vibrational state of the atoms in the trap during the measurement time. Due to the differential ac-Stark shift, this energy change results in a non-reversible broadening of the atomic transition frequency. The resulting temporal reduction of the spin-echo oscillation amplitude can be characterized by the irreversible dephasing time $\mathrm{T_2^\prime}$, defined as the $\mathrm{1/e}$-decay time of the oscillation amplitude \cite{Lengwenus_1}. 
With the application of the compensation laser we measure $\mathrm{T_2^\prime = \unit{(86.4 \pm 4.6)}{ms}}$. This result confirms that our technique of compensation of the differential light shift fully cancels the effects of homogeneous dephasing and pushes the achievable dephasing time to the unavoidable limit given by inhomogeneous dephasing effects.
\\
Finally, we would like to discuss the magnitude of the deteriorating effects of additional spontaneous scattering processes caused by the near-resonant compensation laser field.  Although expected to have a strong negative effect of high scattering rates due to the small detuning which is about half of the hyperfine transition frequency $\mathrm{\omega_{HFS}}$, this is strongly reduced by the small amount of power needed for the compensation field. The ratio of the rates of spontaneous scattering induced by the trapping and compensation laser, respectively, is given by
\begin{equation}
\frac{\Gamma_{sc,c}}{\Gamma_{sc}} \approx \frac{P_c}{P} \left(\frac{2\Delta_{ij}}{\omega_{HFS}}\right)^2 \approx
\left(\frac{\omega_{HFS}}{2\Delta_{ij}}\right)^2 \left(\frac{2\Delta_{ij}}{\omega_{HFS}}\right)^2 = 1
\end{equation} 
Thus, in spite of its small detuning, the compensation light only adds the same rate of spontaneous scattering events as the trapping light induces itself. In addition, the total amount of scattering events can be made arbitrarily small by increasing the detuning of the trapping light. For the parameters of the central trap, we obtain $\mathrm{\Gamma_{sc} \approx \Gamma_{sc,c} \approx 3 s^{-1}}$ \cite{scattering}.\\
%
%
In this work, we presented a universal method for compensation of the differential light shift of pairs of internal states of optically trapped atoms. It utilizes a second light field detuned about halfways between the transition frequencies for coupling both states to the excited state responsible for the trapping potential. With the compensation field, we could minimize the dephasing induced by the differential light shift for atoms in a single trap by a factor of 20 and equalized the transition frequencies in an array of dipole traps of varying  depth to a limit given by the mode-mismatch caused by an imperfect optical setup. The rate of additional scattering events induced is comparable to the one of the trapping light and can be made arbitrarily small by increasing the trap laser detuning. We could demonstrate a significant increase in the dephasing time for thermal atom ensembles in a single trap but also for the ensemble average of many traps. This leads to a synchronisation of the coherent evolution of trapped atomic quantum systems as required for scaling up the number of quantum systems for quantum computing, quantum simulation, and quantum metrology, including advanced atomic clocks.
The straightforward scheme demonstrated is universally applicable and not limited to alkali atoms or the hyperfine state superpositions used for this experimental demonstration and levitates the requirement of a magic-wavelength light field for trapping of atomic quantum systems.

\begin{acknowledgments}
	We acknowledge financial support from the Deutsche Forschungsgemeinschaft (DFG) [Grant No. BI 647/6-1, Priority Program SPP 1929 (GiRyd)].
	M.S. and J.K. contributed equally to this work.
\end{acknowledgments}

\bibliographystyle{apsrev4-1}
\bibliography{comp-lightshift}

\begin{thebibliography}{25}
\bibitem [{r1()}]{r1}%
  \BibitemOpen
  \href@noop {} {}\bibinfo {note} {With the dipole laser at \unit{810.1}{nm}
  the ratio of the differential light shift $\delta_0$ compared to the absolut
  light shift $\Delta$ is $\delta_0/\Delta \approx 3.42\times
  10^{-4}$.}\BibitemShut {Stop}%
\bibitem [{sca()}]{scattering}%
  \BibitemOpen
  \href@noop {} {}\bibinfo {note} {For completeness we would like to note that
  the rate of scattering events inducing heating for the trapping light is
  further reduced for far-detuned traps. This does not alter the given argument
  but only makes the scattering by the trapping light less
  harmful.}\BibitemShut {Stop}%
\end{thebibliography}

\end{document}